\documentclass[times, twocolumn]{aastex701}

\usepackage{amsmath,amssymb,bm,mathtools}
\usepackage{graphicx}
\usepackage{xcolor}
\usepackage[normalem]{ulem}
\usepackage{hyperref}
\usepackage{physics}

\newcommand{\Robs}{R_{\nu}^{\rm obs}}
\newcommand{\Rrest}{R_{\nu}^{\rm rest}}
\newcommand{\Rcore}{R_{\nu}^{\rm core}}
\newcommand{\Rlobe}{R_{\nu}^{\rm lobe}}
\newcommand{\Rsf}{R_{\nu}^{\rm SF}}
\newcommand{\Rjet}{R_{\nu}^{\rm jet}}
\newcommand{\Rexc}{R_{\nu}^{\rm exc}}
\newcommand{\Pjet}{P_{\rm jet}}
\newcommand{\Mdot}{\dot{M}}
\newcommand{\PhiBH}{\Phi_{\rm BH}}
\newcommand{\phibh}{\phi_{\rm BH}}
\newcommand{\Xienv}{\Xi_{\rm env}}
\newcommand{\taunu}{\tau_{\nu}}
\newcommand{\Knu}{\mathcal{K}_{\nu}}
\newcommand{\Gbulk}{\Gamma}
\newcommand{\ahat}{a_\ast}
\newcommand{\Omegah}{\Omega_{\rm H}}

\newcommand{\fduty}{f_{\rm duty}}
\newcommand{\etaj}{\eta_{\mathrm{j}}}
\newcommand{\epssyn}{\epsilon_{\mathrm{syn}}}
\newcommand{\fesc}{f_{\mathrm{esc}}}
\newcommand{\fbeam}{f_{\mathrm{beam}}}

\newcommand{\figplaceholder}[1]{%
  \fbox{\parbox[c][2.25in][c]{0.95\linewidth}{\centering\small #1}}%
}
\newcommand{\incfig}[2][]{%
  \IfFileExists{#2}{\includegraphics[#1]{#2}}{\figplaceholder{Missing figure file: \texttt{#2}}}%
}

\hypersetup{linkcolor=red,citecolor=blue,filecolor=cyan,urlcolor=magenta}

\shorttitle{Time-Domain Radio Loudness}
\shortauthors{An}
\submitjournal{ApJL}

\begin{document}

\title{Time-domain Radio-loudness of Active Galactic Nuclei:\\
       Intermittency, Memory, and Jet Escape}

\author{Tao An}
\affiliation{Department of Astronomy, University of Science and Technology of China, 96 Jinzhai Rd., Hefei, Anhui 230026, China}
\affiliation{Shanghai Astronomical Observatory, Chinese Academy of Sciences,
80 Nandan Road, Shanghai 200030, China}
\email{antao1979@ustc.edu.cn}

\correspondingauthor{Tao An}

\begin{abstract}
The classical radio-loudness parameter $R \equiv f_\nu(5\,\mathrm{GHz})/f_\nu(4400\,\text{\AA})$ divides a radio flux density by an optical/UV accretion tracer, but the two terms do not probe the same clock. The radio numerator can blend compact-core emission from the current engine, lobe and relic plasma left by earlier jet episodes, and host-galaxy synchrotron emission. We introduce a time-domain radio-loudness (TDRL) description that keeps these contributions separate. The radio numerator is written as compact-core and extended-lobe terms, with recovered fractions set by observing frequency, angular resolution, and surface-brightness sensitivity. For a single intermittently jetted AGN population, a two-state duty cycle filtered by exponential lobe fading gives an exact stationary Beta distribution for the normalized extended-radio response. Its mean is $\fduty$, while its variance scales as $(1+\chi_\nu)^{-1}$, where $\chi_\nu\equiv\taunu/t_{\rm switch}$. In this reference limit, the familiar GHz valley near the classical radio-loud/radio-quiet boundary can arise from short radio memory alone, without requiring two intrinsic engine classes. Metre-wave surveys that recover diffuse emission and subtract the host contribution should therefore progressively fill the valley. In the $(\Rcore,\Rlobe)$ plane, a core--lobe mismatch index separates triggering, sustained, and remnant phases, provided orientation-dependent core beaming is modelled or controlled. A complementary two-barrier picture, involving horizon-threading magnetic flux and jet escape through the nuclear medium, separates jet launching from the formation of large-scale radio structure. This view makes radio loudness a probe of jet duty cycle, radio memory, and escape through the host environment.
\end{abstract}


\section{Introduction}

For nearly four decades, $R \equiv f_{\nu}(\mathrm{5\,GHz})/f_{\nu}(4400\,\text{\AA})$
has served as the standard measure of AGN radio loudness
\citep{1989AJ.....98.1195K}.
Whether its observed distribution is genuinely bimodal has been debated since the late 1980s
and remains unsettled; the answer depends critically on sample selection and survey
design \citep{2002AJ....124.2364I,2003MNRAS.346..447C,2007ApJ...658..815S,2016ApJ...831..168K}.
In flux-density-limited GHz-selected samples, sources with large and small $R$ were routinely treated as
two empirically distinct populations separated by a trough in the distribution.
This usage also carried an implicit physical interpretation: objects with small $R$ were often assumed, explicitly or operationally, to lack significant jet activity.

Low-frequency surveys and high-resolution radio imaging have, however, progressively challenged both the empirical validity of this division and the physical picture it implies. Low-frequency interferometric observations that recover diffuse lobe plasma, together with population studies of optically selected quasars, find a broad and largely continuous distribution of radio output \citep{2019A&A...622A..11G,2019A&A...622A..12H,2021MNRAS.506.5888M,2024MNRAS.528.4547A,2023MNRAS.518...39W,2023MNRAS.525.6064W,2024A&A...691A.191C}.
Parsec- and sub-kiloparsec-scale imaging, on the other hand, increasingly uncovers compact cores, jet knots, and mildly relativistic motions in sources that integrated flux catalogues would still classify as radio quiet \citep{2005ApJ...621..123U,2013MNRAS.432.1138P,2022ApJ...936...73A,2022MNRAS.510..718P,
 2023MNRAS.518...39W, 2023MNRAS.525.6064W, 2023MNRAS.525..164C,2025MNRAS.537..705N}.
Independent studies at sub-arcsecond resolution confirm this trend across a range of AGN samples \citep[e.g.,][]{2018MNRAS.476.3478B,2021MNRAS.500.4749B, 2019MNRAS.485.2710J,2021MNRAS.503.1780J}.
This tension arises because scalar $R$ mixes the current central engine state with the cumulative record of earlier jet episodes.
These two quantities need not track each other in any single observation.

The relevant radio components evolve on very different timescales.
Compact core emission traces the central engine on timescales of days to years and is strongly modulated by Doppler boosting.
Extended lobes and relic plasma preserve earlier particle injection and fade only after radiative and dynamical losses act over $10^7$--$10^8$\,yr \citep{1973A&A....26..423J,2007A&A...470..875P}. Arcsecond radio surveys often blend these components, so scalar $R$ mixes present activity with earlier jet history.
The optical/ultraviolet denominator, by contrast, reflects the nuclear luminosity much more nearly at the epoch of observation. Thus a core-bright source may have little extended emission if the jet is young or recently restarted, whereas a source with luminous lobes may now have a weak nucleus.

The long-standing radio-loud/radio-quiet debate can be clarified by recognizing that it
encompasses two separable physical questions.
The first is whether the inner accretion flow can launch a relativistic outflow at all.
The second is whether individual jet episodes are sufficiently powerful and long-lived to
propagate beyond the nuclear interstellar medium and inflate a persistent synchrotron reservoir.
Because these are logically independent conditions, treating them as one has been a principal
source of ambiguity in the literature.
Indeed, both a genuinely two-population picture and a single intermittently cycling population can
generate similar one-dimensional $p(\log R)$ distributions once the compact and extended radio
contributions are blended together.

The apparent dichotomy also depends on how the source is observed.
Varying the radio frequency, angular resolution, or surface-brightness sensitivity changes how
much compact-core emission is isolated, how much diffuse lobe plasma is recovered, and how much
host-galaxy synchrotron radiation remains in the measured flux density.
The effect is strongest near the classical valley: at low frequency the radio numerator is bounded
from below by star-formation and relic plasma, while at high frequency or high resolution the same
source may appear deficient in extended emission.
Radio loudness is therefore more informative when these contributions are kept separate rather than folded into one survey-dependent scalar value.
This perspective aligns naturally with the broader shift from the empirical RL/RQ definition toward a
physically motivated distinction between jetted and non-jetted AGN
\citep{2017NatAs...1E.194P}, and with recent Bayesian decomposition approaches that statistically separate AGN and star-formation contributions to the radio emission of quasars \citep{2024MNRAS.529.3939Y}.

Intermittent AGN jets and long-lived extended radio emission are not new ideas. They appear in spectral-ageing studies \citep{1973A&A....26..423J,2007A&A...470..875P}, dynamical models of remnant and restarted sources \citep{1997ApJ...487L.135R,2020MNRAS.496.1706S}, and analytic or simulation-based population models that couple duty cycles to spectral ageing and source evolution \citep{2013ApJ...769..129S,2015ApJ...806...59T,2018MNRAS.475.2768H}. Radio morphology statistics, including LoTSS census work, also provide independent constraints on radio-AGN duty cycles and host-property correlations \citep{2019MNRAS.488.2701M}. The classical core-dominance parameter has long separated beamed and unbeamed components \citep[e.g.,][]{1995PASP..107..803U}. Our aim is to put these elements into one analytic description in which the timescale mismatch is explicit and testable. 

We develop a time-domain description of radio loudness with three components (Section~\ref{sec:framework}). First, we derive a closed-form stationary distribution for the extended-radio response in a minimal intermittent-jet model. Second, we define the observable pair $(\Rcore,\Rlobe)$, which separates the prompt engine state from the time-integrated jet history; the accumulated lobe emission retains the effects of frequency, resolution, surface-brightness selection, and host-galaxy radio emission. Third, we relate the decomposition to a two-barrier picture in which jet launching and large-scale escape are distinct physical requirements. Section \ref{sec:phenomenology} shows how the framework recasts the RL/RQ problem in terms of intermittency, radio memory, and propagation through the ambient medium.It then discusses its relationship between this framework and the accretion-mode taxonomies of radio AGN. Section~\ref{sec:applications} develops the main observational tests and survey strategy and
Section~\ref{sec:conclusions} provides conclusions.

\section{Time-Domain Radio-Loudness Framework}\label{sec:framework}
The decomposition below defines the time-domain radio-loudness (TDRL) framework used throughout the paper.

\subsection{Observable definitions}
\label{subsec:defR}

The historical radio-loudness definition is an observed flux-density ratio,
\begin{equation}
\Robs(t)\ \equiv\ \frac{f_{\nu_{\rm r}}^{\rm obs}(t)}{f_{\nu_{\rm opt}}^{\rm obs}(t)}\,,
\label{eq:Robs}
\end{equation}
with $\nu_{\rm r}=5$\,GHz and $\nu_{\rm opt}=c/4400\,\text{\AA}$ in the
classical convention \citep{1989AJ.....98.1195K}.
For physical interpretation it is convenient to work instead with the rest-frame
specific-luminosity ratio,
\begin{equation}
\Rrest(t)\ \equiv\ \frac{L_{\nu_{\rm r}}^{\rm rest}(t)}{L_{\nu_{\rm opt}}^{\rm rest}(t)}\,,
\label{eq:Rrest}
\end{equation}
where both luminosities are evaluated at the same rest-frame frequencies.
For power-law spectra $f_\nu\propto\nu^\alpha$ the two are related by
\begin{equation}
\Rrest = \Robs\,(1+z)^{\alpha_{\rm r}-\alpha_{\rm opt}}\,,
\label{eq:Kcorr}
\end{equation}
so any RL/RQ comparison spanning a broad redshift range depends implicitly on the adopted
K-correction unless this is made explicit.
In what follows we use $\Robs$ when discussing survey measurements and $\Rrest$ when
connecting to intrinsic jet power.
We note that for sources near the RL/RQ boundary, the optical spectrum may be dominated by host starlight rather than a power-law AGN continuum, invalidating the simple K-correction of Eq.~(\ref{eq:Kcorr}). In such cases a host-subtracted optical luminosity or an X-ray normalisation may be more appropriate.

\subsection{Decomposing the radio numerator}\label{subsec:decomp}

A measured radio flux density generally combines several physically distinct components.
At fixed observing frequency and angular resolution, the total is
\begin{equation}
f_{\nu_{\rm r}}^{\rm obs}(t)
= w_{\rm c}\,f_{\nu_{\rm r}}^{\rm core}(t)
+ w_{\rm l}\,f_{\nu_{\rm r}}^{\rm lobe}(t)
+ w_{\rm SF}\,f_{\nu_{\rm r}}^{\rm SF}(t)\,,
\label{eq:radio_decomp}
\end{equation}
where $f_{\nu_{\rm r}}^{\rm core}$ is compact, beaming-sensitive emission from the jet base;
$f_{\nu_{\rm r}}^{\rm lobe}$ is optically thin extended emission from lobes, relics, and diffuse
structures; and $f_{\nu_{\rm r}}^{\rm SF}$ is host-galaxy radio emission dominated by star formation.
The coefficients $w_i\in[0,1]$ are observational recovery factors. They depend on angular resolution, surface-brightness depth, $uv$ coverage, and the model used to separate compact and diffuse emission.
Differences in reported radio-loud fractions between surveys of the same parent population are
therefore not necessarily inconsistent; they reflect different weighted sums of the same
underlying flux components, sampled through different observational filters.
In practice, the $w_i$ are not purely instrumental: they depend on the assumed source model used in image decomposition and on how extended and compact components are separated, which introduces a model-dependent element into the decomposition.

We replace the scalar $R$ with the two-component quantity 
\begin{equation} \bm{\mathcal{R}}_\nu(t)\ \equiv\ \left(\Rcore(t),\,\Rlobe(t)\right),
\label{eq:TDRL_vector}
\end{equation}
where $\Rcore\equiv f_{\nu_{\rm r}}^{\rm core}/f_{\nu_{\rm opt}}$ and
$\Rlobe\equiv f_{\nu_{\rm r}}^{\rm lobe}/f_{\nu_{\rm opt}}$,
with consistent K-corrections applied to all flux densities. These quantities are observational radio-to-optical ratios, not pure jet-power estimators. Their physical interpretation as core and lobe diagnostics requires the optical/UV denominator to be controlled, measured contemporaneously, or replaced by an alternative accretion proxy when denominator variability or host contamination is important.

\begin{figure}[t]
\centering
\incfig[width=\columnwidth]{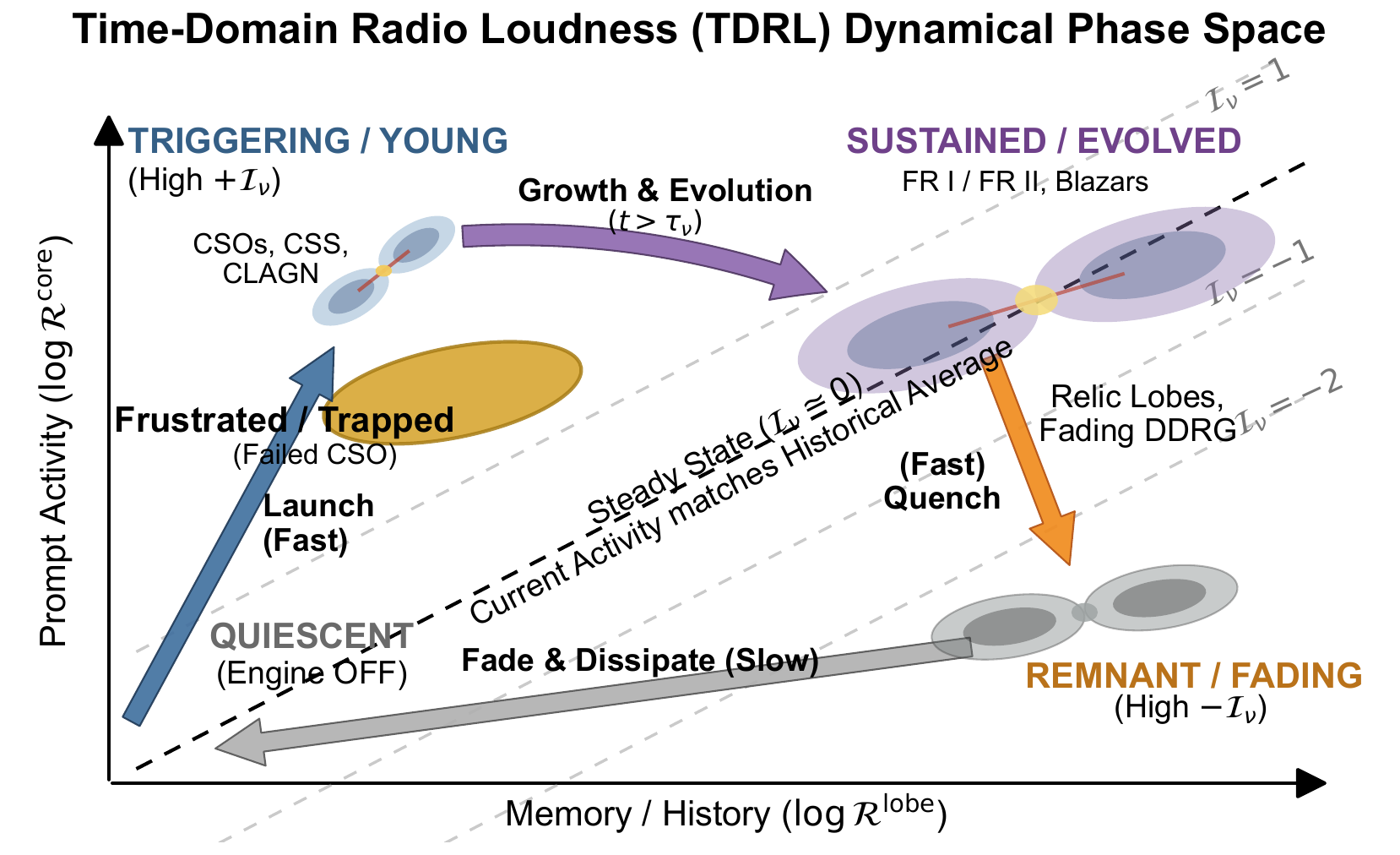}
\caption{The TDRL diagnostic plane. The vertical axis shows the prompt/core radio loudness $\Rcore$, tracing present-day engine activity; the horizontal axis shows the memory/lobe radio loudness $\Rlobe$, tracing the time-integrated jet history over $\taunu$. The dashed diagonal marks approximate quasi-equilibrium between injection and accumulated lobe power; the perpendicular offset defines the mismatch index $\mathcal{I}_\nu$ (Eq.~\ref{eq:mismatch}). Four regimes are indicated: triggering or restarting (core bright, lobe faint, $\mathcal{I}_\nu\gg0$), sustained activity (both components bright, $\mathcal{I}_\nu\simeq0$), remnant or fading (lobe bright, core faint, $\mathcal{I}_\nu\ll0$), and quiescent or weak systems (both components faint). The two-dimensional decomposition separates what the engine is doing now from what it has done in the past, a distinction that scalar $R$ cannot make. Orientation-dependent beaming of the core component can shift sources along the vertical axis and must be modelled or controlled for in quantitative applications.}
\label{fig:tdrl}
\end{figure}

We quantify the core--lobe mismatch index with
\begin{equation}
\mathcal{I}_\nu\ \equiv\ \log_{10}\!\left[\frac{\Rcore}{\Rlobe+\epsilon}\right]\quad(\mathrm{dex}),
\label{eq:mismatch}
\end{equation}
where $\epsilon$ is a lobe-detection floor, expressed in the same radio-to-optical units as $\Rlobe$.
Equivalently, $\epsilon=f_{\nu_{\rm r}}^{\rm lobe,lim}/f_{\nu_{\rm opt}}$ for an adopted upper limit to undetected extended emission.
It regularizes the ratio when $\Rlobe$ is consistent with noise, but it is not fixed uniquely by the image rms.
A surface-brightness threshold must be converted to a total-flux-density limit by assuming, or marginalizing over, the angular size and surface-brightness profile of the missing lobe.
As shown in Figure \ref{fig:tdrl}, large positive $\mathcal{I}_\nu$ identifies core-dominated, lobe-faint systems: jets that
are newly triggered, recently restarted, or still confined within the circumnuclear medium.
Large negative $\mathcal{I}_\nu$ identifies remnant systems in which the central engine has
faded while fossil lobe plasma remains detectable at low frequency.
$\mathcal{I}_\nu \approx 0$ marks the quasi-steady regime, where the current injection
luminosity and the accumulated lobe energy are in approximate balance.
This is not meant to imply a universal physical steady state.
Because compact-core and lobe spectra differ, the location and interpretation of $\mathcal{I}_\nu=0$ are frequency dependent and must be defined for a specified observing setup.
When using survey data, $\epsilon$ should therefore be treated as a survey- and source-model-dependent censoring parameter; different choices will shift the $\mathcal{I}_\nu$ distribution, so inter-survey comparisons must adopt a common convention.

The mismatch index $\mathcal{I}_\nu$ is closely related to the classical core dominance parameter $R_{\rm c}\equiv S_{\rm core}/S_{\rm ext}$ \citep[e.g.][]{1995PASP..107..803U}, which has long been used to separate beamed from unbeamed components.
Indeed, if the same optical denominator is used for the core, lobe, and lobe upper-limit terms, the denominator cancels in Eq.~(\ref{eq:mismatch}).
In that limit $\mathcal{I}_\nu$ is simply $\log_{10}R_{\rm c}$ with an explicit upper-limit regularization.
Its purpose is therefore not to define a new orientation-free core-dominance measure, but to embed core prominence in the TDRL bookkeeping of frequency, resolution, surface-brightness selection, upper limits, and host-galaxy subtraction.

\paragraph{Star-formation floor and jet-excess loudness.}
The host galaxy star-formation term $f_{\nu_{\rm r}}^{\rm SF}$ is a predictable baseline, and it must be estimated before the radio numerator can be interpreted as a jet diagnostic. Over Myr--Gyr timescales, host synchrotron emission is quasi-steady relative to the AGN optical/UV continuum and can dominate the radio flux density at low frequency (a few hundred MHz) when the jet is inactive. We therefore define a jet-excess radio loudness,
\begin{equation}
\Rexc(t)\ \equiv\ \frac{f_{\nu_{\rm r}}^{\rm obs}(t) - w_{\rm SF}\,f_{\nu_{\rm r}}^{\rm SF,pred}}
                        {f_{\nu_{\rm opt}}^{\rm obs}(t)}
          \ \simeq\ w_{\rm c}\Rcore(t) + w_{\rm l}\Rlobe(t)\,,
\label{eq:Rexcess}
\end{equation}
where $f_{\nu_{\rm r}}^{\rm SF,pred}$ is estimated from host-galaxy star-formation modelling.
In practice this should preferably be done with panchromatic SED fitting, followed by a radio luminosity--SFR calibration matched to the observing frequency \citep[e.g.,][]{2018MNRAS.475.3010G}.
Individual tracers such as infrared luminosity, UV+IR, or recombination lines remain useful, but each can be biased in radiatively efficient AGN if used in isolation \citep{1992ARA&A..30..575C,2001ApJ...554..803Y,2003ApJ...586..794B}.
The distinction between $\Robs$ and $\Rexc$ matters most at the low-$R$ end, where the classical RL/RQ valley is most sensitive to host galaxy contamination.

\subsection{Intermittent engines and frequency-dependent radio memory}\label{subsec:memory}

Two assumptions are sufficient for the time-domain argument: jet power varies, and extended radio emission remembers earlier injection. To make the intermittency concrete we anchor it to a specific physical mechanism, though the formalism itself is independent of this choice. In GRMHD simulations, the magnetic flux threading the black-hole event horizon, $\phi_{\rm BH}$, is consistently identified as the primary control parameter for jet efficiency \citep{2011MNRAS.418L..79T}. This flux evolves through a competition among field advection from the outer disc, turbulent diffusion, and reconnection, none of which attains a steady state. Coherent field loops arrive stochastically from the outer disc and circumgalactic environment, driving $\phibh$ on a random walk about a saturation threshold $\phi_{\rm BH,crit}$; threshold crossings produce sporadic strong-jet episodes superposed on a background of weak or quiescent activity (Appendix \ref{app:flux}). EHT horizon-scale polarimetric imaging of M87$^\ast$ has confirmed that dynamically important ordered magnetic fields exist close to the horizon \citep{2021ApJ...910L..13E}, supporting the plausibility of dynamically important near-horizon magnetic fields.

\begin{figure*}[t]
\centering
\incfig[width=0.9\textwidth]{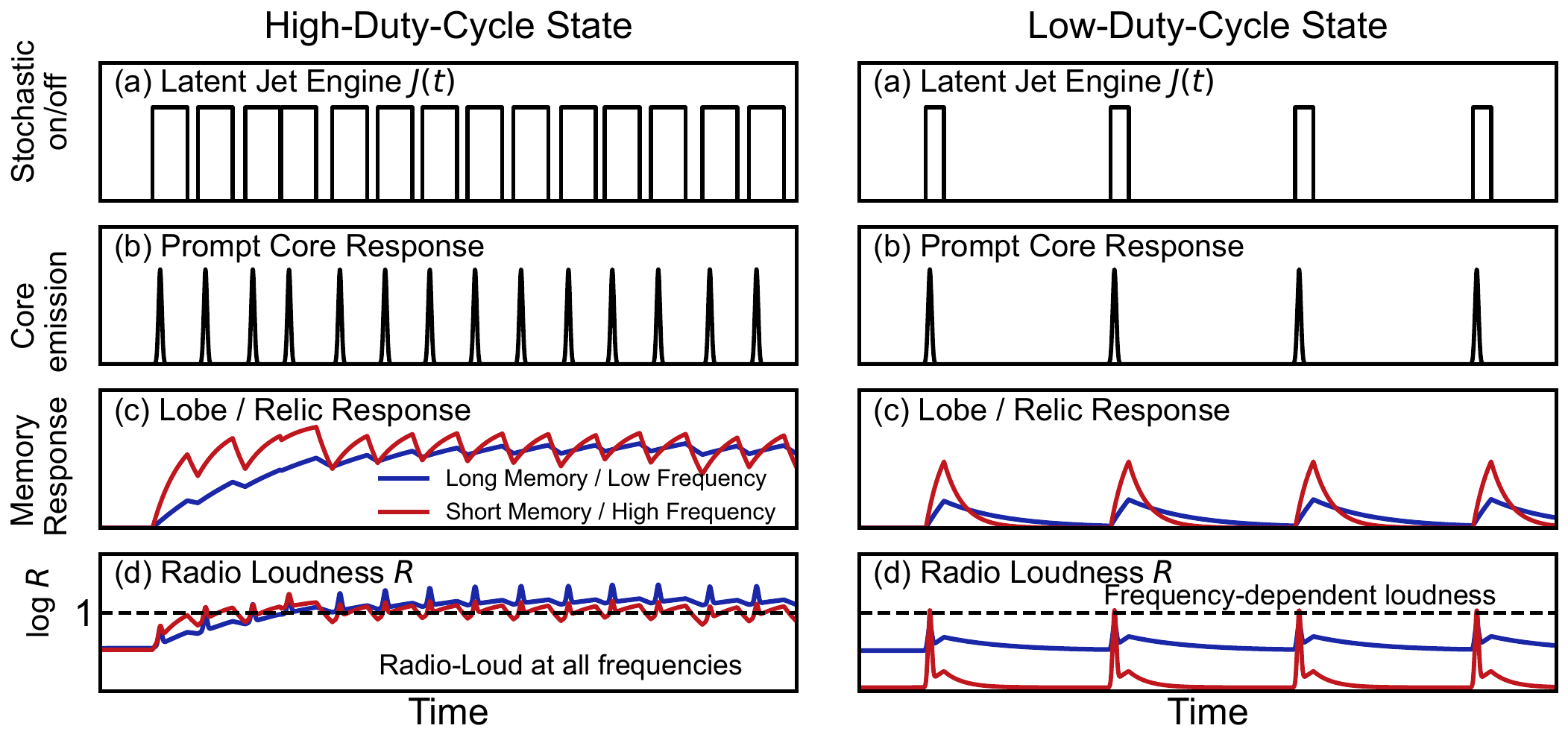}
\caption{Time-domain origin of radio loudness in the TDRL framework. Panels show: (a) the underlying engine state $J(t)$ modelled as a stochastic on/off process, (b) the prompt compact-core emission modulated by beaming, (c) the lobe emission smoothed by the frequency-dependent fading function for short-memory (high frequency, green) and long-memory (low frequency, red) observations, and (d) the resulting observed scalar radio loudness $R$. A single intermittent engine produces either a bimodal or a continuous $p(\log R)$ depending on $\chi_\nu=\taunu/t_{\rm switch}$ and on whether the survey recovers diffuse emission.}
\label{fig:telegraph}
\end{figure*}

On timescales long compared with individual threshold crossings, a broad class of such stochastic processes is well represented by a two-state random telegraph process $J(t)\in\{0,1\}$ with transition rates $\lambda_\uparrow$ (off$\to$on) and $\lambda_\downarrow$ (on$\to$off) (Figure \ref{fig:telegraph}). The  duty cycle and mean episode durations are then
\begin{equation}
\fduty\ \equiv\ \langle J\rangle
       = \frac{\lambda_\uparrow}{\lambda_\uparrow+\lambda_\downarrow},\quad
\langle t_{\rm off}\rangle = \lambda_\uparrow^{-1},\quad
\langle t_{\rm on}\rangle  = \lambda_\downarrow^{-1}\,.
\label{eq:fduty}
\end{equation}

The imprint of this switching is carried most directly by the extended lobe luminosity rather than by the integrated loudness ratio. We therefore write
\begin{equation}
L_{\nu}^{\rm lobe}(t)\ \propto\ \int_{-\infty}^{t} dt'\,\Knu(t-t')\,J(t')\,,
\end{equation}
where $\Knu$ is the fading function describing how previously injected plasma contributes to the observed emission at frequency $\nu$. After division by an optical denominator that varies more slowly than the radio numerator, or after conditioning on a narrow optical-luminosity range, the same convolution structure is inherited by $\Rlobe$. The explicit form of $\Knu$, including the exponential approximation adopted for analytic work, is given in Appendix~\ref{app:function}.

The effective memory time $\taunu$ includes transport, adiabatic losses, and flux blending. Its frequency dependence is nevertheless constrained by synchrotron and inverse-Compton cooling.  For synchrotron and inverse-Compton losses in a lobe field $B$ with CMB-equivalent field $B_{\rm CMB}\simeq3.25(1+z)^2\,\mu{\rm G}$, the radiative lifetime of electrons emitting primarily at rest-frame frequency $\nu$ is
\begin{equation}
t_{\rm rad}(\nu)\ \simeq\ 1590\,{\rm Myr}\,
\frac{B^{1/2}}{B^2+B_{\rm CMB}^2}\, \nu^{-1/2},
\label{eq:trad}
\end{equation}
consistent with classical spectral-ageing models \citep{1973A&A....26..423J,2007A&A...470..875P}. Here $B$ is in $\mu$G, $\nu$ is the rest-frame frequency in GHz, and $t_{\rm rad}$ is in Myr. The numerical coefficient follows from the synchrotron characteristic frequency $\nu_c = (3/2)\gamma^2 eB/(2\pi m_e c)$ and the energy-loss rate $d\gamma/dt = -(4/3)\sigma_T c \gamma^2 (B^2+B_{\rm CMB}^2)/(8\pi m_e c^2)$, evaluated at the frequency where $\nu_c = \nu$. At fixed lobe conditions, $\taunu\propto\nu^{-1/2}$, with additional shortening at high redshift where $B_{\rm CMB}^2$ dominates the total loss rate \citep{2015MNRAS.452.3457G,2017MNRAS.468..109W}. This $\nu^{-1/2}$ scaling is the origin of the frequency dependence in the TDRL framework: the memory time is not a free parameter to be fitted survey by survey but a quantity with a known frequency and redshift dependence that can be predicted from lobe conditions and tested against spectral curvature data.

\subsection{Two separable barriers: launching and escape}\label{subsec:barriers}

Building a large-scale radio source requires sustained jet activity over $10^7$--$10^8$\,yr, a condition that compact radio sources need not satisfy, and separating the two regimes is essential for interpreting the observed radio-loudness distribution. The \emph{launching barrier} is governed by the horizon-threading magnetic flux, expressed as $\phibh\propto\PhiBH/\sqrt{\Mdot c}$, which separates low-flux SANE-like states from high-flux MAD-like states in GRMHD phenomenology \citep{1977MNRAS.179..433B,2003PASJ...55L..69N,2011MNRAS.418L..79T}. The \textit{escape barrier} is set by whether a jet episode is sufficiently long-lived, and sufficiently powerful, for the jet head to break out of the dense, clumpy circumnuclear medium; otherwise the source remains compact or fades before reaching large scales \citep{1997ApJ...487L.135R,2012ApJ...760...77A,2016MNRAS.461..967M,2018MNRAS.475.3493B}.
The escape term is not required for the time-domain mechanism itself. A broad distribution of jet lifetimes, biased toward short episodes, can by itself produce many more compact than large sources.
We introduce $\Xienv$ as a phenomenological way to connect lifetime statistics to propagation through the ambient medium, not as evidence for a unique frustrated-source channel.
In many elliptical hosts the hot gas atmosphere may dominate the source dynamics from kpc to Mpc scales, while cold or clumpy gas is an additional nuclear-scale complication in some systems.
We define the escape parameter as the ratio
\begin{equation}
\Xienv\ \equiv\ \frac{t_{\rm on}}{t_{\rm bo}}\,,
\label{eq:Xi_def}
\end{equation}
where $t_{\rm bo}$ is the time for the jet head to traverse the relevant gas scale height (Appendix~\ref{app:escape}). Episodes with $\Xienv\gtrsim1$ inflate large-scale lobes; episodes with $\Xienv\lesssim1$ deposit their kinetic energy within the nuclear region and produce no extended emission  (Figure~\ref{fig:phase}).

In practice the escape transition is gradual rather than sharp. A convenient smooth form is
\begin{equation}
\fesc(\Xienv)\ =\ S\!\left(\frac{\Xienv-1}{\sigma_\Xi}\right),
\quad S(x)\equiv \frac{1}{1+e^{-x}},
\label{eq:fesc_eq}
\end{equation}
where $\sigma_\Xi$ captures source-to-source scatter in gas geometry and clumpiness.
In the limit $\sigma_\Xi\to0$ this reduces to the sharp step at $\Xienv=1$.

Jet history feeds back on the escape condition in a physically consequential way. Previous episodes can excavate low-density passages through the nuclear gas, lowering the effective ambient density encountered by subsequent jets. This decreases $t_{\rm bo}$ and raises $\Xienv$ even if the engine statistics are unchanged, providing a natural route by which repeated compact episodes eventually culminate in large-scale breakout and linking Compact Symmetric Objects (CSOs) and Compact Steep Spectrum (CSS) phenomenology to jet intermittency without invoking permanent confinement \citep{2012ApJ...760...77A}. In population calculations this feedback can be captured by letting $t_{\rm bo}$ depend on the cumulative injected energy in the inner kiloparsec, an effect potentially accessible through cold-gas and ionized-outflow diagnostics.

\subsection{Stationary distribution of the extended radio component}\label{subsec:analytic_solution}

An exact solution is available in a useful limiting case. Suppose the jet alternates between a strong state that feeds the extended radio source at an approximately constant rate and a weak state whose contribution to the large-scale emission is negligible. 

We work with the radio numerator directly rather than with the full loudness ratio, because the numerator is where the memory resides. The normalized extended-radio response is
\begin{equation}
Y(t)\equiv \frac{L_{\nu}^{\rm lobe}(t)}{L_{\nu,{\rm max}}^{\rm lobe}},
\label{eq:Y_def_main}
\end{equation}
where $L_{\nu,{\rm max}}^{\rm lobe}$ is the asymptotic lobe luminosity of a source that remains
continuously in the strong state at the same rest-frame frequency.
This quantity is only the normalization of the exponential-response reference model.
It should not be read as the physical asymptote of a real lobe, whose luminosity can rise to a maximum and then decline as dynamical expansion and radiative losses evolve with environment and redshift \citep[e.g.,][]{1997MNRAS.292..723K,2012ApJ...760...77A,2013MNRAS.430..174H,2014MNRAS.443.1482H,2018MNRAS.475.2768H,2019MNRAS.490.5807E}.
With the exponential response function (Appendix~\ref{app:function}), $Y$ obeys
\begin{equation}
\taunu\,\frac{dY}{dt}\ =\ J(t) - Y(t)\,,
\label{eq:Y_ode_main}
\end{equation}
where $J(t)$ is the telegraph process introduced above.
The derivation in Appendix~\ref{app:beta_deriv} then gives the stationary probability density
\begin{equation}
p(Y) = \frac{\Gamma(\alpha+\beta)}{\Gamma(\alpha)\,\Gamma(\beta)}\,
       Y^{\alpha-1}(1-Y)^{\beta-1},
\label{eq:beta_dist}
\end{equation}
with shape parameters
\begin{equation}
\alpha = \lambda_\uparrow\taunu,\qquad \beta = \lambda_\downarrow\taunu\,.
\label{eq:alpha_beta}
\end{equation}

The physical content of this result is concentrated in its first two moments:
\begin{align}
\mathbb{E}[Y] &= \frac{\alpha}{\alpha+\beta} = \fduty, \label{eq:meanY}\\[4pt]
{\rm Var}(Y)  &= \frac{\fduty(1-\fduty)}{1+\chi_\nu}, \label{eq:varY}
\end{align}
where
\begin{equation}
\chi_\nu\ \equiv\ (\lambda_\uparrow+\lambda_\downarrow)\taunu
         = \frac{\taunu}{t_{\rm switch}},\quad
t_{\rm switch}\equiv \frac{1}{\lambda_\uparrow+\lambda_\downarrow}.
\label{eq:chi}
\end{equation}
The first two moments show what the model can and cannot measure. The mean traces the duty cycle. The variance traces the ratio of fading time to switching time. 
The distribution piles up toward $Y\simeq0$ and $Y\simeq1$ only when $\alpha<1$ and $\beta<1$, i.e.\ when the radio fading time is shorter than both the mean ON and OFF episode durations. At GHz frequencies sources could plausibly be in this regime; at metre wavelengths they generally are not, helping to explain why the RL/RQ valley appears prominent in some surveys and dissolves in others.

The Beta form is explicit for the two-state, exponential-response model and should be regarded as the minimal analytic reference solution. More realistic fading laws or a distribution of jet powers will modify the detailed shape of $p(Y)$, but the same physical competition remains: the duty cycle sets the mean extended emission, and the ratio $\taunu/t_{\rm switch}$ sets how sharply the population clusters about that mean. When sources are compared within narrow optical-luminosity bins, or when optical variability is subdominant to the dynamic range of the radio numerator, the same scaling is inherited by $\Rlobe$.

Because compact and extended emission can often be measured separately, the conditional stationary distributions are also analytically accessible:
\begin{equation}
p(Y\,|\,J=1)\propto Y^\alpha(1-Y)^{\beta-1},\qquad
p(Y\,|\,J=0)\propto Y^{\alpha-1}(1-Y)^\beta\,,
\label{eq:condY}
\end{equation}
corresponding to Beta$(\alpha+1,\beta)$ and Beta$(\alpha,\beta+1)$ after normalization. These forms yield closed predictions for the fractions of recently triggered and remnant systems once the survey sensitivity, resolution, and source selection are specified (Appendix~\ref{app:forward}).

\paragraph{Robustness of the analytic solution.}
The Beta distribution is exact only within the two-state, exponential-response limit. Three idealizations are most relevant for interpreting it as a reference solution.

First, real jets exhibit a continuous distribution of kinetic powers rather than a binary on/off state. Convolving the telegraph process with a log-normal power distribution would broaden the resulting $p(Y)$ and soften any bimodal peaks. Second, the exponential fading function is an effective one-timescale approximation. More realistic spectral-ageing models \citep[Jaffe--Perola, KP;][]{1973A&A....26..423J} produce sharper high-frequency cutoffs that would steepen the tails of $p(Y)$ at high $\nu$ while leaving the low-frequency behaviour less affected. Third, a multi-state Markov chain with intermediate jet efficiencies would smooth the boundary between the $J=0$ and $J=1$ conditional distributions.

These extensions would change the detailed shape of $p(Y)$, but not the basic expectation that short memory favours bimodality and long memory suppresses it. The quantitative variance scaling $(1+\chi_\nu)^{-1}$ will depend on the assumed power distribution and fading law. The Beta form should therefore be regarded as a minimal reference solution, while precise parameter inference from observed distributions will require population models that relax these idealizations.

\subsection{Illustrative population model}\label{subsec:forward_demo}

We next give a minimal population calculation to show the scale of the effect for observationally plausible parameters. The calculation is not fitted to any survey. Its purpose is to illustrate the predicted frequency dependence of the extended-radio variance and the resulting change in valley contrast.
The physically relevant prediction is therefore the frequency-dependent change in the extended-radio variance, rather than the exact location or depth of the illustrative GHz valley. Moderate changes in $\fduty$, $t_{\rm switch}$, or the adopted host-galaxy floor would shift the peak positions and valley contrast, but would not change the qualitative expectation that shorter radio memory produces a broader, more valley-prone distribution than longer radio memory.

We adopt a duty cycle $\fduty=0.10$, consistent with the $\sim$5--15\% radio-loud fraction observed in optically selected quasar samples \citep{2016ApJ...831..168K}. We note that equating $\fduty$ with the observed radio-loud fraction is only approximately valid, because the radio-loud fraction is itself defined through the scalar $R$ that TDRL seeks to replace. The correspondence holds to the extent that, in the minimal model, the mean of the extended-radio distribution equals $\fduty$ (Eq.~\ref{eq:meanY}), and the radio-loud fraction at GHz frequencies is set primarily by the high-$Y$ tail. For the illustrative purpose here, this approximation is sufficient.

We adopt a mean switching timescale $t_{\rm switch}=5$\,Myr, which falls within the broad range spanned by independent constraints: dynamical and spectral ages of CSO and CSS sources ($\sim10^2$--$10^5$\,yr; \citealt{1999A&A...345..769M, 2012ApJ...760...77A, 2021A&ARv..29....3O, 2024ApJ...961..240K, 2024ApJ...961..242R}), spectral-ageing estimates of remnant and restarted radio galaxies ($\sim10^7$--$10^8$\,yr cycle times; \citealt{2020MNRAS.496.1706S}), and GRMHD-motivated magnetic-flux evolution timescales. The adopted value is intended to illustrate the regime where $\chi_\nu$ transitions from order unity at GHz frequencies to $\chi_\nu\gg1$ at metre wavelengths; the behaviour of the model is insensitive to the precise choice within a factor of a few. For the radio memory we use Eq.~(\ref{eq:trad}) with a representative lobe field $B=10\,\mu$G at $z=0.5$, giving $\taunu\simeq 70$\,Myr at 150\,MHz ($\chi_\nu\approx14$) and $\taunu\simeq 12$\,Myr at 5\,GHz ($\chi_\nu\approx2.4$). The normalized extended-radio variable $Y$ is drawn from the Beta distribution of Eq.~(\ref{eq:beta_dist}) with the corresponding $\alpha=\lambda_\uparrow\taunu$ and $\beta=\lambda_\downarrow\taunu$. A log-normal host star-formation floor with mean $\log R_{\rm SF}=-0.3$ and scatter 0.3\,dex is added to produce the total observed $\log R$.

\begin{figure*}[t]
\centering
\incfig[width=\textwidth]{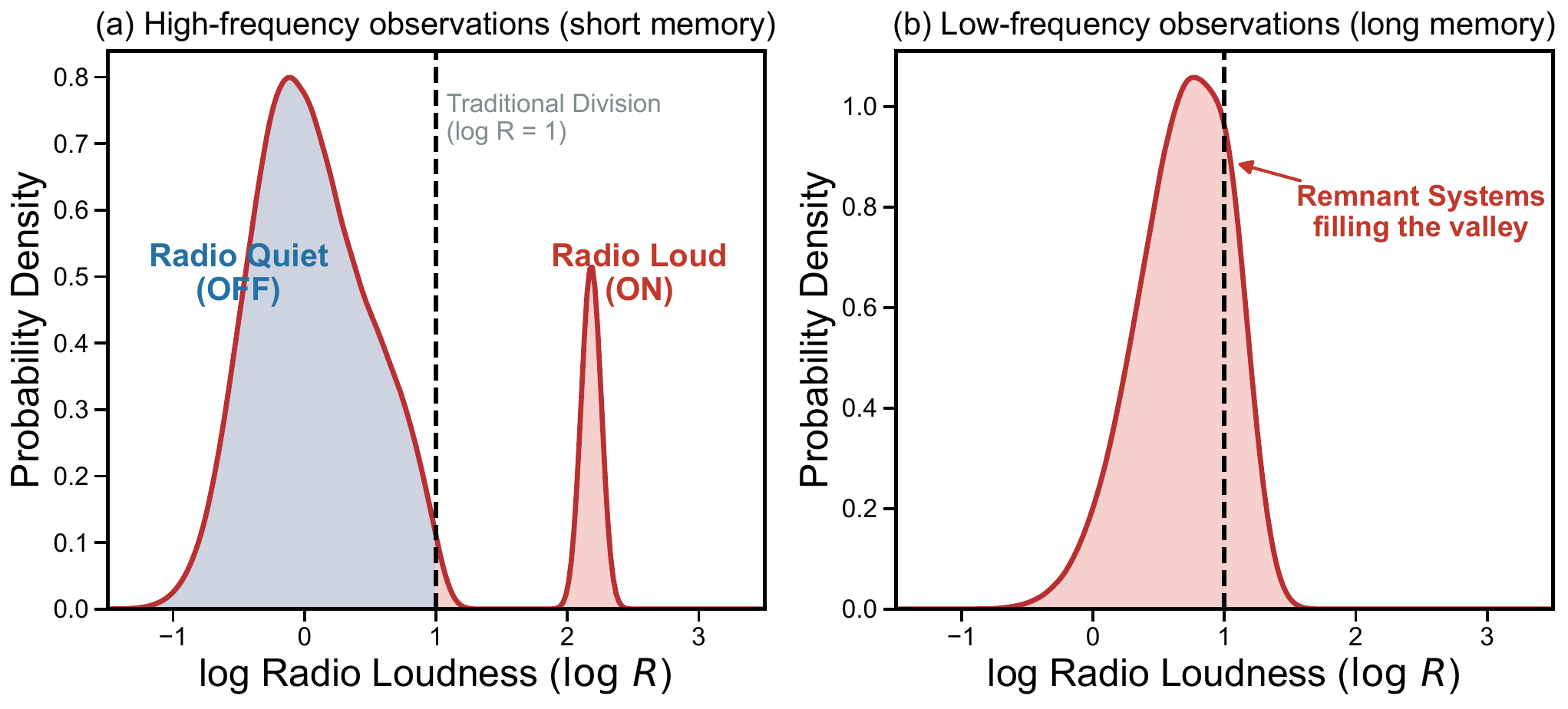}
\caption{Illustrative frequency-dependent radio-loudness distributions
from a single intermittent engine population
($f_{\rm duty}\simeq 10\%$). Both panels adopt the illustrative
parameters of Section~2.6 ($t_{\rm switch}=5\,$Myr, $B=10\,\mu$G,
$z=0.5$) with a log-normal host star-formation floor
($\langle\log R_{\rm SF}\rangle=-0.3$, scatter $0.3\,$dex).
(a)~At high frequencies ($\nu_{\rm obs}=5\,$GHz,
$\chi_\nu\approx 2.4$, short memory) the Beta shape parameter
$\alpha=\lambda_\uparrow\tau_\nu<1$, so the stationary distribution
of the normalized extended-radio response piles up near $Y\approx 0$;
combined with the host star-formation floor this produces a bimodal
$p(\log R)$ with a pronounced valley near the classical
$\log R=1$ boundary. (b)~At low frequencies
($\nu_{\rm obs}=150\,$MHz, $\chi_\nu\approx 14$, long memory)
$\alpha>1$ and the Beta distribution becomes unimodal; fading and
remnant systems fill the valley, yielding a broad, continuous
distribution. The variance of the normalized extended-radio response
shrinks as $(1+\chi_\nu)^{-1}$ (Eq.~\ref{eq:varY}). These distributions are an existence proof, not a fit to data. The adopted parameters are illustrative (Section~\ref{subsec:forward_demo}). Proper interpretation of
observed samples requires subtracting the host contribution and
accounting for surface-brightness selection.}
\label{fig:forward}
\end{figure*}

At 5\,GHz ($\chi_\nu\approx2.4$, short memory), the resulting $p(\log R)$ is bimodal with a valley near $\log R\simeq1$, in general agreement with classical GHz samples  (Figure \ref{fig:forward}). At 150\,MHz ($\chi_\nu\approx14$, long memory), the variance of the extended component shrinks by a factor $\sim$5 and the distribution becomes broad and unimodal, consistent with the continuous distributions reported from LoTSS-based studies \citep{2019A&A...622A..11G,2021MNRAS.506.5888M,2019A&A...622A..12H}. We reiterate that this experiment does not constitute a fit to any specific survey sample. It confirms only that the TDRL mechanism can produce the generally observed phenomenology with plausible parameter values. Whether this mechanism is the dominant driver of the observed RL/RQ distribution, or merely one contributor among others, requires the quantitative multi-frequency comparison deferred to future work.

\section{Phenomenology of the TDRL Framework}\label{sec:phenomenology}

We now trace how the TDRL decomposition bears on several long-standing issues in AGN radio phenomenology. In each case, the difficulty comes from that compact and extended radio emission record jet activity on timescales differing by orders of magnitude. Ignoring that difference leads to apparent contradictions.

\paragraph{The RL/RQ valley as the outcome of finite radio memory plus a host contribution.}
The stationary Beta distribution already contains the essential physics of the valley (Section \ref{subsec:analytic_solution}). For a single parent population, short-memory GHz measurements sample sources close to their instantaneous states and can therefore appear bimodal, whereas deep metre-wave observations average over many switching cycles and yield a broader, more continuous distribution. 
The most robust prediction is not the presence or absence of a valley, which depends strongly on sample construction. It is the frequency dependence of the normalized extended-radio variance. 
In the minimal reference model, ${\rm Var}(Y)$ is predicted to fall toward low frequency as $(1+\chi_\nu)^{-1}$ (Eq.~\ref{eq:varY}), and any valley in the distribution should fill once $\chi_\nu\gtrsim1$. Redshift provides a second lever on the same physics. At fixed observed frequency, a higher-redshift source is observed at a higher rest-frame radio frequency, and its lobe electrons cool more rapidly because $U_{\rm CMB}\propto(1+z)^4$. The effective memory time therefore shortens toward the early Universe, so even deep metre-wave observations do not automatically provide a long-memory view at high $z$. This reverses a common expectation: radio quietness at high redshift constitutes \emph{weaker}, not stronger, evidence against jet production than it does locally. A jet may well be launched, yet inverse-Compton losses can erode the diffuse synchrotron reservoir before it ever becomes prominent in the radio band \citep{2015MNRAS.452.3457G,2017MNRAS.468..109W}. The $z=7$ blazar recently reported   fits this picture \citep{2025NatAs...9..293B}: the core is visible; the lobes are not.

At low $R$, the observed radio numerator is often dominated by star-formation emission and faint relic plasma even when the jet is off. Two effects conspire to shape the low-$R$ peak in $p(\log R)$. First, $\Robs$ carries a floor and scatter set by the host SFR and the radio--infrared correlation \citep{1992ARA&A..30..575C,2001ApJ...554..803Y}, so the low-$R$ peak in $p(\log R)$ is not a pure jet diagnostic. Second, the depth of the valley between the classical RL and RQ peaks is particularly sensitive to whether old lobes are recovered ($w_{\rm l}\simeq1$) and whether the host contribution is removed (i.e.\ whether $\Robs$ is replaced by $\Rexc$). The quantity most directly tied to jet physics is therefore $p(\log\Rexc)$ after correcting for resolution and surface-brightness selection.

\paragraph{The core--lobe mismatch as a phase diagnostic.}
A scalar $R$ conflates whether a jet is currently being launched with whether earlier episodes built an extended synchrotron reservoir. The pair $(\Rcore,\Rlobe)$ and the mismatch index $\mathcal{I}_\nu$ keep these two pieces of information separate, distributing sources across a diagnostic plane rather than collapsing them onto a single locus (Figure~\ref{fig:tdrl}). Sources trace hysteresis loops in this plane as the core responds promptly to state changes while the lobe term accumulates and decays on the longer timescale $\taunu$. Large positive mismatch identifies triggering or restarting systems in which the compact source is already bright but the extended structure has not yet built up or has been confined by the local environment. Large negative mismatch identifies remnant systems in which the present-day engine has faded but fossil lobes persist. Near-zero mismatch marks the quasi-steady regime of sustained injection.

\paragraph{Orientation effects on the diagnostic plane.}
Because $\Rcore$ includes Doppler-boosted emission from the jet base (Eq.~\ref{eq:Lcore}), sources viewed close to the jet axis will have systematically elevated $\Rcore$ at fixed intrinsic jet power. A blazar, for example, may appear in the triggering quadrant (high $\Rcore$, low $\Rlobe$) purely because of orientation rather than genuinely recent jet onset. Conversely, a source viewed at large angles may have suppressed $\Rcore$ and occupy the remnant quadrant even during an active phase.
Lobe-dominated or steep-spectrum selection does not remove this problem.
In such samples the core can be Doppler-suppressed rather than merely unbeamed, so the observed $\Rcore$ may span several dex at fixed intrinsic jet power.
For parsec-scale Lorentz factors of order $\Gamma\sim10$, orientation alone can introduce a four-dex dispersion in core prominence, comparable to that seen in complete flux-limited samples \citep[e.g.,][]{2009MNRAS.398.1989M}.
The diagnostic plane is therefore most useful as a statistical framework after orientation indicators, sidedness, variability, spectral shape, or an explicit beaming model $\fbeam(\theta,\Gamma)$ have been included.
This degeneracy does not invalidate the diagnostic plane, but it limits source-by-source evolutionary classification and must be accounted for in any quantitative population study.

\begin{figure}[t]
\centering
\incfig[width=\columnwidth]{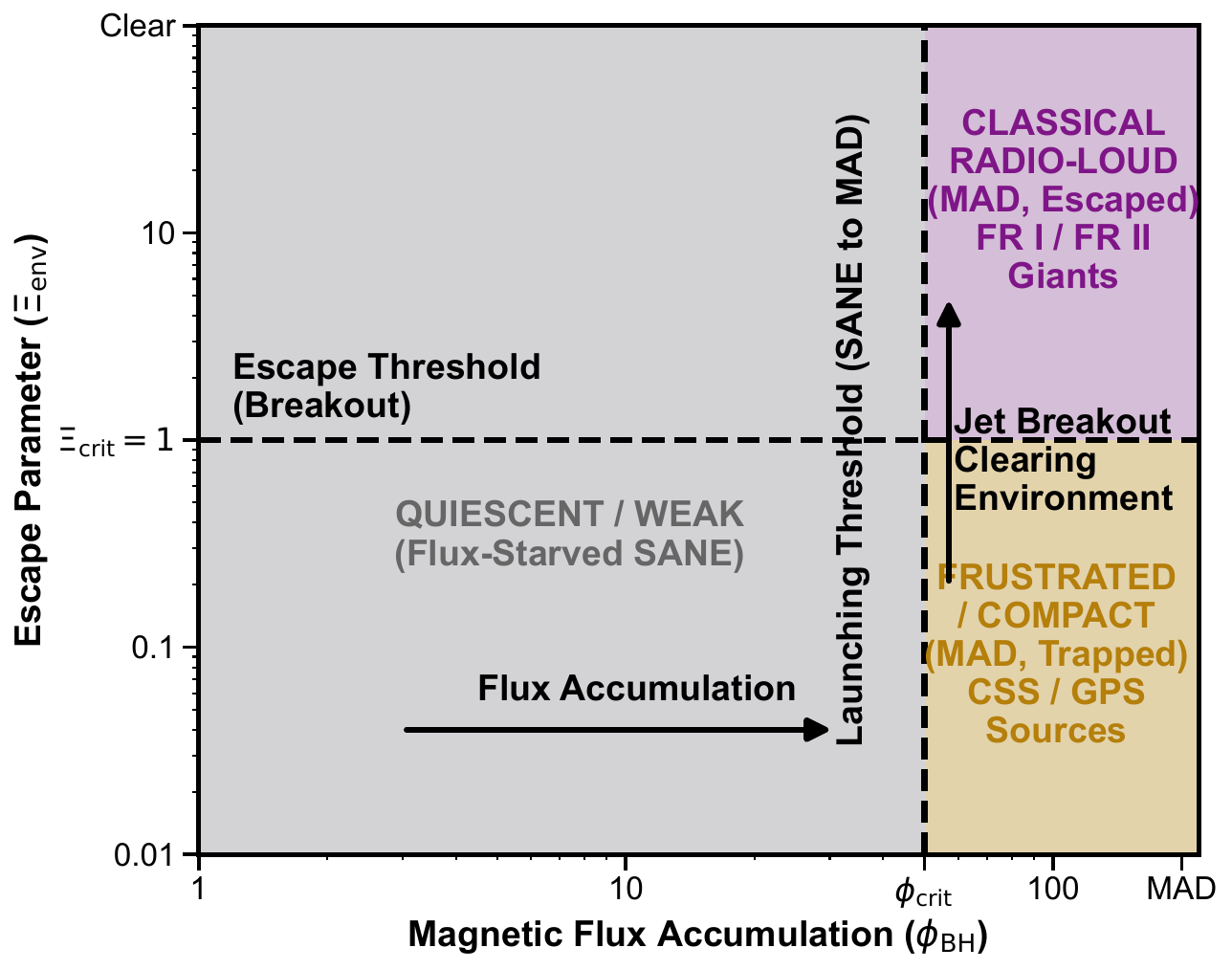}
\caption{Heuristic two-barrier phase diagram for AGN jet lifecycles. The horizontal axis shows the dimensionless horizon-threading magnetic flux $\phibh$ (SANE$\rightarrow$MAD); the vertical axis shows the escape parameter $\Xienv=t_{\rm on}/t_{\rm bo}$, separating confined from escaping jets. Three regimes are schematically indicated: quiescent or weak systems (flux-starved SANE), frustrated or compact sources such as CSO/CSS objects (MAD jets confined by dense environments, $\Xienv\lesssim1$), and classical large-scale radio galaxies (MAD jets that escape and inflate lobes, $\Xienv\gtrsim1$). Engine power and environmental permeability are independent axes: a jet can be powerful yet remain compact if the circumnuclear gas column is sufficiently high. The diagram classifies known phenomenology into physically motivated regions but does not predict population fractions (see Section~\ref{sec:phenomenology} for discussion).}
\label{fig:phase}
\end{figure}

\paragraph{Launching and escape are separable.}
In the two-barrier $(\phibh,\Xienv)$ plane (Figure~\ref{fig:phase}), three physically distinct regimes can be distinguished. At low $\phibh$, flux-starved flows produce only weak, wind-like outflows that dissipate rapidly and build no significant lobe reservoir. At high $\phibh$ but low $\Xienv$, the central engine launches a powerful outflow, yet dense nuclear gas enforces strong entrainment and rapid decollimation, thermalizing the jet on sub-kpc scales; these are the compact, core-bright sources associated with CSO/CSS phenomenology \citep{1998PASP..110..493O,2012ApJ...760...77A,2021A&ARv..29....3O}. When both barriers are cleared (high $\phibh$, high $\Xienv$), jets inflate large-scale lobes in classical extended radio galaxies \citep[e.g.,][]{1974MNRAS.166..513S, 1974MNRAS.169..395B, 1989ApJ...345L..21B, 2020NewAR..8801539H}, whose low-frequency emission provides an approximately isotropic, time-integrated measure of past jet power and of the kinetic feedback exerted on the surrounding medium \citep{2007ARA&A..45..117M,2012ARA&A..50..455F}.

We emphasise that the $(\phibh,\Xienv)$ plane as presented in Figure~\ref{fig:phase} is a heuristic organizing scheme rather than a predictive population model. $\phibh$ requires horizon-scale magnetic-flux measurements currently available only for M87$^\ast$ and a few other sources, while $\Xienv$ depends on the circumnuclear gas density, scale height, and jet cross-section. All these vary by orders of magnitude across the population and are accessible only through indirect diagnostics, such as cold-gas mass and ionized-outflow kinematics. The diagram classifies known phenomenology into physically motivated regions but does not, in its current form, predict the relative populations in each quadrant as a function of observable quantities such as black-hole mass, Eddington ratio, or host-galaxy gas fraction. Converting the two-barrier picture into  population predictions will require coupling the TDRL framework to semi-analytic models of magnetic-flux evolution and circumnuclear gas dynamics. That step is outside the scope of this paper. 

\paragraph{Relationship to the spin paradigm and two-population models.}
The TDRL framework demonstrates that a single intermittently jetted population \emph{can} reproduce the RL/RQ valley, but it does not mean that a two-population model is excluded.
On theoretical grounds, black-hole spin and horizon-threading magnetic flux are expected to influence jet efficiency \citep{1977MNRAS.179..433B,2003PASJ...55L..69N,2011MNRAS.418L..79T}. However, a secure observed correlation between radio loudness and spin has not yet been established \citep[e.g.,][]{2007ApJ...658..815S,2013A&A...557L...7V}, largely because direct spin constraints are available for only a small number of AGN and remain observationally demanding \citep{2021ARA&A..59..117R}.
Independent lines of evidence nevertheless suggest that host-galaxy properties may set the \emph{probability} of entering a high-duty-cycle, high-$\phi_{\rm BH}$ state, including the strong association of powerful radio jets with giant elliptical hosts and merger histories \citep{2005MNRAS.362....9B, 2015ApJ...806..147C}, and the relative rarity of powerful jets in disc-dominated systems \citep{2022A&A...662A..20V, 2022A&A...668A..91V, 2026arXiv260222668R}. The TDRL framework is compatible with this picture: a distribution of $\fduty$ and $\Xi_{\rm env}$ across the population, correlated with spin and host properties, would produce a composite $p(\log R)$ that could resemble either a single broad distribution or a bimodal one depending on the sharpness of the underlying parameter correlations. Distinguishing the two scenarios requires model comparison in the full $(\Rcore,\Rlobe)$ plane, conditioned on black-hole mass, Eddington ratio, and host morphology, rather than one-dimensional $p(\log R)$ alone.
The comparison with accretion-mode taxonomies also requires care \citep{2012MNRAS.421.1569B,2014ARA&A..52..589H,2020NewAR..8801539H}.
The present $(\Rcore,\Rlobe)$ plane is normalized by the radiative output and therefore applies most directly to radiatively efficient AGN with a measurable optical/UV denominator \citep{2014ARA&A..52..589H}.
For radiatively inefficient jet-mode AGN, there is no equally direct measure of the instantaneous accretion rate in this normalization, because hot accretion flows are weakly radiative and often coupled to jet/outflow power rather than to a thin-disc optical/UV continuum \citep{2008ARA&A..46..475H,2014ARA&A..52..529Y}.
Accretion-mode differences should therefore be tested by conditioning the TDRL variables on independent mode indicators, or by adopting an alternative denominator, rather than by assuming that different modes must occupy distinct loci in the same plane \citep{2003MNRAS.345.1057M,2004A&A...414..895F,2012MNRAS.421.1569B}.

\paragraph{Relativistic cores in nominally radio-quiet AGN.}
Compact radio cores in nominally radio-quiet AGN are not surprising in a time-domain picture. The growing number of VLBI detections of compact cores, mildly relativistic motions, and sub-kiloparsec jet features in nominally radio-quiet objects \citep{2022ApJ...936...73A,2022MNRAS.510..718P, 2023MNRAS.525..164C,2021MNRAS.504.3823W,2023ApJ...944..187W,2023MNRAS.523L..30W,2023MNRAS.518...39W,2023MNRAS.525.6064W,2025ApJ...987L..26W} is consistent with the TDRL picture, which places such sources in the triggering or restarting region of the $(\Rcore,\Rlobe)$ plane. We note, however, that this expectation is not unique to TDRL: any framework that allows episodic jet activity would predict occasional compact detections. The specific added value of the TDRL decomposition is the prediction that these sources should cluster at high $\mathcal{I}_\nu$ (top-left corner in the $(\Rcore,\Rlobe)$ plane) and that their lobe emission should follow the Beta distribution conditioned on $J=1$ (Eq.~\ref{eq:condY}). These sources are more naturally interpreted as young or recently restarted radio sources than as systems intrinsically devoid of large-scale jets. The e-MERLIN Quasar Feedback Survey makes this especially clear: once the inner $\sim10^2$\,pc are resolved, AGN-related radio structures appear in the majority of a predominantly radio-quiet quasar sample \citep{2025MNRAS.537..705N}. Scalar $R$ misses them, while the mismatch index does not.

A necessary caveat is that not every compact radio source is a jet. Recent parsec-scale work on radio-quiet quasars suggests that some compact sources are consistent with self-absorbed synchrotron from very small scales, whereas others may include wind-shock or coronal contributions \citep{2023MNRAS.525..164C,2024ApJ...975...35C,2025ApJ...979..241C, 2025NatAs...9.1086L}. In practical applications, $\Rcore$ should therefore be read as the prompt compact AGN contribution; when a genuinely jet-specific subsample is required, brightness temperature, polarization, flat or inverted spectra, variability, and proper motions provide the most reliable discriminants \citep{2023MNRAS.525.6064W}.

\paragraph{Remnant radio galaxies.}
Deep low-frequency imaging has revealed substantial populations of remnant and restarted radio galaxies \citep{2017A&A...606A..98B,2019A&A...622A..13M,2020MNRAS.496.1706S,2023Galax..11...74M,2024Galax..12...11M}. In the TDRL framework these occupy the low-$\Rcore$/high-$\Rlobe$ quadrant, reflecting the inertia of the extended synchrotron reservoir after jet cessation. A direct prediction is that a non-negligible fraction of optically selected, apparently radio-quiet AGN should be surrounded by faint, spectrally aged structures invisible to high-resolution GHz surveys but recoverable in surface-brightness-limited metre-wave maps. Measuring that fraction would constrain the duty-cycle distribution convolved with $\taunu$.

\paragraph{Changing-look AGN and non-simultaneous radio loudness.}
In changing-look events the optical denominator of $R$ can vary by orders of magnitude on year-like timescales \citep{2015ApJ...800..144L}. Non-simultaneous radio and optical measurements therefore generate apparent excursions in scalar $R$ even if the radio source itself is steady. TDRL predicts a distinctive trajectory in the $(\Rcore,\Rlobe)$ plane: on such timescales the lobe term is frozen while $\Rcore$ may respond with a finite lag. Changing-look AGN should therefore trace strongly non-circular trajectories in the $(\Rcore,\Rlobe)$ plane rather than merely crossing the historical $R=10$ threshold. Specifically, consider a turn-off event in which the optical luminosity drops by a factor $\delta_{\rm opt}\sim10$--100 over $\Delta t\sim1$--10\,yr. Because $\taunu\gg\Delta t$, the lobe term $\Rlobe$ increases by $\sim\delta_{\rm opt}$ (the denominator shrinks while the numerator is frozen), producing a near-vertical upward excursion in the $(\Rcore,\Rlobe)$ plane. The core term $\Rcore$ may track the optical decline with a lag of order the light-travel time across the jet base ($\sim$weeks to months), producing a brief clockwise loop. In a subsequent turn-on event the trajectory reverses. The predicted signature is therefore a rapid, predominantly vertical oscillation in $\log\Rlobe$ at nearly fixed $\log\Rcore$ on year-like timescales, distinguishable from genuine jet-state changes that move sources along the diagonal.

\section{Observational Tests}\label{sec:applications}

Testing TDRL requires the same parent sample to have three kinds of measurements:
(i) low-frequency imaging with sufficient surface-brightness sensitivity to recover the extended
emission and approximate $\Rlobe$;
(ii) high-frequency or high-resolution imaging to isolate $\Rcore$ and control beaming; and
(iii) host-galaxy information, together with a principled treatment of upper limits, to estimate
the star-formation contribution.
Current facilities already permit well-defined pilot studies.
LoTSS and VLASS provide an existing low- and higher-frequency wide-area pairing, while LSST/Rubin will improve the optical variability and host-galaxy side of the problem \citep{2019A&A...622A...1S,2020PASP..132c5001L,2019ApJ...873..111I}.
In the longer term, SKA-Low, SKA-MID, and ngVLA will test the key predictions at high statistical significance \citep[e.g.,][]{2015aska.confE..86N}.

The present paper develops the analytic reference framework and identifies its testable predictions.
Here we outline the key discriminants and the survey strategy required.

\subsection{Key observational discriminants}\label{subsec:decidable}

The central question is whether a second, intrinsically distinct engine class is required, or whether intermittency within a single population is enough to reproduce the observed diversity. In TDRL, the compact core approximates the present engine state, while the diffuse extended emission integrates past jet activity over the fading timescale $\taunu$. The two components constrain different epochs of the same engine.

\paragraph{Test 1: population in the mismatch quadrants of $(\Rcore,\Rlobe)$.}

A single intermittent population predicts substantial numbers of triggering systems (core bright, lobe faint) and remnant systems (core faint, lobe bright), producing broad and often asymmetric tails in $\mathcal{I}_\nu$. A strict two-population model predicts fewer such sources once orientation and measurement scatter are accounted for. Comparison between a two-component mixture and a single switching model convolved with the radio response time can be performed directly on matched samples.

\paragraph{Test 2: frequency dependence of the extended-emission distribution.}

Intermittency predicts that the width of the extended-radio distribution narrows toward lower frequency following Eqs.~(\ref{eq:varY}) and (\ref{eq:chi}), and that the mismatch distribution correlates with spectral-curvature and spectral-ageing indicators. A static two-population model may exhibit some survey dependence. However, it does not naturally produce a variance scaling as $(1+\chi_\nu)^{-1}$ when the same parent sample is observed at different frequencies.

\paragraph{Test 3: redshift dependence of radio memory.}
The low-frequency trend should weaken, or partially reverse, toward high redshift as inverse-Compton losses against the CMB shorten $\taunu$. The memory-based interpretation predicts a clear redshift evolution. If the remnant fraction or extended-emission variance remains constant across matched samples, this framework is directly challenged.

\paragraph{Test 4: consistency between compact-core variability and extended emission.}
Multi-epoch GHz monitoring, supplemented by VLBI, constrains the switching timescale and duty cycle of the compact sources. These parameters predict the expected distribution of extended emission through $\chi_\nu\equiv\taunu/t_{\rm switch}$. The core-inferred duty cycle must therefore align with the observed lobe distribution of the parent sample. An incompatibility between them would indicate that the compact and extended terms lack a common switching history.

\subsection{From catalogue radio loudness to TDRL observables}\label{subsec:implementation}

In practice, $\Rcore$ and $\Rlobe$ must be measured for the same optically or X-ray selected parent sample. Constructing such a sample from heterogeneous radio catalogues observed at different frequencies and resolutions is precisely the conflation that TDRL is designed to resolve. A compact detection does not automatically imply a jet. Population studies allow two approaches. One can retain $\Rcore$ as the full prompt compact AGN term. This treats non-jet compact emission as an additional contribution. Alternatively, one can restrict the analysis strictly to jets. This requires filtering for high brightness temperature, clear collimated morphology, flat or inverted spectra, or measurable proper motions. The first route preserves completeness, while the second yields a smaller but more secure jet-selected sample. Either approach is valid provided the sample definition is applied consistently through the population analysis.

Samples should be drawn from optically or X-ray selected AGN with well-characterized completeness and ancillary infrared photometry for star-formation subtraction. For each source, a core-weighted measurement at rest-frame $\gtrsim5$--$15$\,GHz or at sub-arcsecond resolution isolates the nucleus. A lobe-weighted measurement at $\sim50$--$350$\,MHz with $uv$-tapering to preserve diffuse emission constrains the extended term. In addition, independent SFR tracers supply $f_\nu^{\rm SF,pred}$. With these in hand, $\Rcore$, $\Rlobe$, $\Rexc$, and $\mathcal{I}_\nu$ are defined source by source rather than merged into a single flux-density ratio. Existing facilities already make well-defined pilot samples feasible: LOFAR/LoTSS provides deep low-frequency imaging for large optical samples in the northern hemisphere \citep{2019A&A...622A...1S}; VLASS already provides wide-area 2--4\,GHz imaging over most of the VLA-visible sky and is currently the most practical high-frequency counterpart to LoTSS for large samples \citep{2020PASP..132c5001L}. Targeted VLA, e-MERLIN, and VLBI observations can then isolate compact emission for selected subsamples, rather than serving as all-sky $>5$\,GHz surveys. MeerKAT and Australian Square Kilometre Array Pathfinder (ASKAP) are building the southern counterpart. Upper limits on $\Rlobe$ are not missing data but constraints on the weak or off state and must be incorporated through survival analysis or hierarchical inference.

\subsection{Survey strategy with current and future facilities}\label{subsec:ska}

SKA-Low will deliver metre-wave measurements of diffuse emission at the surface-brightness depth needed to detect faint relic structures around optically selected AGN and to map spectral curvature across lobes, directly constraining $\taunu(\nu)$. SKA-MID will provide the higher-frequency and better-matched sky coverage needed to connect those low-frequency measurements to compact and intermediate-scale emission. ngVLA, supplemented by VLBI, will isolate compact cores with the sensitivity required to characterize present-day jet activity well into the radio-quiet population. Controlling the denominator is equally important. Sample matching in redshift, black-hole mass, and Eddington ratio is essential, since variations in the optical denominator and host contribution can otherwise mimic changes in jet duty cycle. LSST/Rubin will be valuable here because repeated optical photometry helps separate genuine radio-state changes from variability in the denominator. The cadence requirement is strongly asymmetric. Extended emission evolves only on long timescales and can generally be measured once per source per survey epoch, whereas the compact core should be revisited on month-to-year baselines to constrain $t_{\rm switch}$ and separate genuine state changes from stochastic variability. The most informative survey designs are therefore those that pair deep low-frequency imaging with repeated high-frequency or high-resolution monitoring of the same optically selected sample.

\subsection{Population inference with explicit survey selection}\label{subsec:stats}

$\Rcore$ and $\Rlobe$ enter catalogues through the selection weights $w_i$ (Section \ref{subsec:decomp}), and  non-detections carry information about the jet-off state (Section \ref{subsec:implementation}). 
The population analysis should be hierarchical, because the detections and upper limits depend on the survey weights (Appendix~\ref{app:forward}). A hierarchical model treats $(\fduty, \, t_{\rm switch}, \, \taunu, \, \Xienv)$ as population-level parameters, allows them to vary across the sample, and predicts the joint likelihood of the observed $(\Rcore,\Rlobe,\Rsf)$ including upper limits. In this formulation, SKA-Low, SKA-MID, and ngVLA are particularly powerful. They reduce censoring by providing complementary low- and high-frequency constraints. Together, these measurements constrain the same underlying duty-cycle history.

\section{Conclusions}\label{sec:conclusions}

The radio-loud/radio-quiet debate has persisted for four decades. The difficulty has not been the absence of relevant physics, but the compression of several timescales into one number. The central observable $R$ conflates processes across six orders of magnitude in timescale. The classical radio numerator does not trace a single engine state. It is a survey-dependent mixture of prompt compact emission, delayed lobe and relic emission with its own fading time, and a host-galaxy contribution. Once that decomposition is made explicit, several strands of apparently contradictory phenomenology can be placed in a common context. Our principal results are as follows.

\begin{enumerate}
    \item \textit{The classical parameter $R$ is not an intrinsic source property unless the observing setup is specified.}    Frequency, angular resolution, surface-brightness sensitivity, and host subtraction each alter the mixture of $\Rcore$, $\Rlobe$, and $\Rsf$ recovered by a given survey, and hence the numerical value of $R$.

    \item \textit{For an intermittent jet with an exponential response function, the normalized extended-radio component follows a stationary Beta distribution whose mean and variance encode distinct physical processes.}
    The mean $\langle Y\rangle$ is determined solely by the jet duty cycle $\fduty$. The variance ${\rm Var}(Y)=\fduty(1-\fduty)/(1+\chi_\nu)$ encodes the competition between radio fading and engine switching. This separation of roles between mean and variance is the central analytic result of the population distribution.

    \item \textit{In the minimal reference model, the classical RL/RQ valley can arise as a short-memory phenomenon and does not, by itself, require two distinct engine species.}
    A two-population contribution is not excluded. However, the valley alone provides insufficient evidence for such an engine dichotomy. Distinguishing between the one- and two-population scenarios requires model comparison in the full $(\Rcore,\Rlobe)$ plane conditioned on black-hole mass, Eddington ratio, and host morphology. As diffuse emission is recovered and the host contribution removed, the valley should weaken. Because inverse-Compton losses against the CMB suppress the extended synchrotron reservoir and shorten the effective lobe memory, radio quietness becomes a less reliable indicator of jet absence at high redshifts than it is locally.

    \item \textit{The pair $(\Rcore,\Rlobe)$ and the mismatch index $\mathcal{I}_\nu$ carry substantially more physical information than scalar $R$.} 
    They separate the triggering, sustained, and remnant phases of jet activity. The extended-emission diagnostics are less sensitive to spurious excursions driven by optical variability in the denominator. Orientation-dependent beaming of the core component is a significant source of contamination in the diagnostic plane and must be modelled or controlled for in population studies.

    \item \textit{Jet phenomenology can be organised around two separable barriers.} 
    A launching barrier controlled by the horizon-threading magnetic flux $\phibh$, and an escape barrier $\Xienv$ set by whether a jet episode persists long enough to break out of the nuclear medium. CSO/CSS sources arise naturally when the first condition is met but the second is not. Repeated failed episodes can excavate the nuclear environment, eventually enabling large-scale breakout without any fundamental change in engine properties. This two-barrier picture is a qualitative organising scheme. Translating it into quantitative population predictions will require coupling to models of magnetic-flux evolution and circumnuclear gas dynamics.

    \item \textit{Compact radio emission in nominally radio-quiet AGN is a natural outcome of this framework rather than an anomaly.}
    These sources reside in the triggering region of the $(\Rcore,\Rlobe)$ plane. As prompt compact emission is not always a pure jet tracer, jet-specific inference benefits from brightness temperature, polarization, spectral slope, variability, or proper-motion diagnostics rather than from compactness alone.

    \item \textit{The predictions of TDRL can be tested with current and near-future facilities.}
    Deep metre-wave imaging measures the long-lived extended term, and high-resolution GHz or VLBI monitoring isolates the compact term. Pairing the two for optically selected samples with controlled host subtraction provides the most informative test.
\end{enumerate}

The TDRL framework moves beyond the scalar $R$ by treating radio loudness as the sum of two physically distinct channels: a compact component tied most closely to the current jet state, and an extended component that records past jet activity over a frequency-dependent fading time. A single-frequency catalogue therefore does not isolate a single physical variable. It mixes how often the engine launches a jet, how long the resulting plasma remains visible, and whether each episode breaks out to form large-scale emission.
This reframes the RL/RQ problem as a question about histories rather than labels. The relevant observables are the duty cycle, the radio-memory time, and the breakout probability, together with their dependence on black-hole mass, accretion rate, environment, and redshift. The present framework does not yet determine whether quasars share one broad duty-cycle distribution or divide into sub-populations with distinct characteristic duty cycles.
Measuring these quantities would connect short jet episodes in individual AGN to the time-integrated kinetic energy supplied by the population, and hence to the cumulative role of jets in circumgalactic gas and galaxy evolution.

\begin{acknowledgments}

I am grateful to the annonymous reviewer for constructive comments and suggestions. I thank Ken Kellermann for careful reading and insightful comments.
\end{acknowledgments}

\appendix
\renewcommand{\theHequation}{\thesection.\arabic{equation}}

The main text draws on three analytic results derived in the appendices:
(i) jet power can be idealized as a two-state process characterized by a duty cycle $\fduty$
and a switching timescale $t_{\rm switch}$;
(ii) extended radio emission is a convolution of that power history with a
frequency-dependent fading function characterized by $\taunu$; and
(iii) the apparent RL/RQ topology depends primarily on the ratio $\chi_\nu\equiv\taunu/t_{\rm switch}$.
The appendices are organized as follows. 
Appendix~\ref{app:flux} presents one stochastic realization of jet switching. Appendix~\ref{app:function} defines the compact and extended response terms. Appendix~\ref{app:escape} gives the escape-barrier scalings. Appendix~\ref{app:beta_deriv} derives the stationary Beta distribution. Appendix~\ref{app:forward} connects the latent duty cycle to observed loudness distributions.

\section{Stochastic magnetic-flux dynamics and jet intermittency}\label{app:flux}

We use magnetic-flux evolution as a concrete realization of the intermittent jet state $J(t)$. Magnetic flux near the horizon executes stochastic excursions, and jet efficiency rises sharply once $\phibh$ crosses a saturation threshold. In GRMHD simulations, jet efficiency is controlled primarily by the dimensionless magnetic flux threading the event horizon \citep{2011MNRAS.418L..79T},
\begin{equation}
\phibh\ \equiv\ \frac{\PhiBH}{\sqrt{\dot{M}\,r_{\rm g}^2\,c}},
\qquad r_{\rm g}\ \equiv\ \frac{GM_\bullet}{c^2},
\label{eq:phibh}
\end{equation}
where $\PhiBH$ is the horizon-threading magnetic flux, $\dot{M}$ the accretion rate,
and $M_\bullet$ the black-hole mass.
The jet power extracted via the Blandford--Znajek mechanism \citep{1977MNRAS.179..433B} is
\begin{equation}
\Pjet\ \simeq\ \frac{\kappa}{4\pi c}\,\PhiBH^2\,\Omegah^2\,f(\Omegah),
\label{eq:Pjet}
\end{equation}
where $\kappa$ depends on field geometry, $\Omegah$ is the horizon angular frequency, and $f(\Omegah)$ captures high-spin corrections \citep{2011MNRAS.418L..79T}. Two accretion regimes are well established: (1)~SANE (Standard and Normal Evolution), with sub-saturation flux $\phibh\ll\phi_{\rm BH,crit}$ and low jet efficiency; (2)~MAD (Magnetically Arrested Disk), with saturated flux $\phibh\approx\phi_{\rm BH,crit}$ and high-efficiency Blandford--Znajek jets. The saturation threshold $\phi_{\rm BH,crit}$ is a phenomenological quantity with weak dependence on field geometry and disc thickness.

We model the jet-efficiency transition as a logistic function of $\phibh$,
\begin{equation}
\etaj(\phibh,\ahat)\ \approx\ \etaj^{\rm w}
+ \left(\etaj^{\rm s}-\etaj^{\rm w}\right)\,
  S\!\left(\frac{\log\phibh-\log\phi_{\rm BH,crit}}{\sigma_{\phibh}}\right),
\label{eq:eta_logistic}
\end{equation}
where $S(x)\equiv(1+e^{-x})^{-1}$, $\etaj^{\rm s}$ and $\etaj^{\rm w}$ are the strong-jet (MAD) and weak-jet (SANE) efficiencies, and $\sigma_{\phibh}$ controls the transition width. Modest fluctuations in the magnetic flux supply can therefore drive rapid transitions between jet states, generating bimodality in jet power from the accretion dynamics alone without invoking two distinct black-hole populations.

The magnetic flux threading the horizon reflects a competition among supply, advection, diffusion, reconnection, and large-scale field topology, none of which need be stationary. Because coherent field loops arrive stochastically from the outer disc, the relevant phenomenology is captured by the minimal stochastic equation
\begin{equation}
\frac{d\PhiBH}{dt}
= \dot{\Phi}_{\rm sup}(t) - \frac{\PhiBH}{t_{\rm diff}} + \sigma_\Phi\,W(t),
\label{eq:flux_sde}
\end{equation}
where $\dot{\Phi}_{\rm sup}$ denotes intermittent flux supply, $t_{\rm diff}$ is an effective flux diffusion and leakage, and $W(t)$ represents short-timescale turbulent fluctuations. The dissipation term $\PhiBH/t_{\rm diff}$ captures the systematic, mean-field loss of coherent flux through reconnection and outward diffusion on the viscous timescale of the inner disc, while the stochastic term $\sigma_\Phi W(t)$ represents short-timescale turbulent fluctuations in the local flux transport that are not captured by the mean-field decay. These are physically distinct processes operating on different timescales: the former sets the equilibrium flux level in the absence of fresh supply, while the latter drives excursions about that equilibrium. Fed through the sharp efficiency transition of Eq.~(\ref{eq:eta_logistic}), this stochastic flux naturally produces sporadic strong-jet episodes as $\phibh$ wanders above $\phi_{\rm BH,crit}$. On timescales much longer than individual transition events, a broad class of such processes reduces to the effective two-state telegraph used in the main text, with switching rates $\lambda_\uparrow$ and $\lambda_\downarrow$ that depend on the flux supply statistics. The horizon-scale polarimetric results for M87$^\ast$ demonstrate that ordered near-horizon magnetic structures strong enough to matter dynamically are astrophysically realized \citep{2021ApJ...910L..13E}.

\section{Extended-radio response and the exponential approximation}\label{app:function}
This appendix states the response model used in the analytic calculation and makes its assumptions explicit. 
The TDRL decomposition of radio loudness into three terms with distinct temporal responses is
\begin{equation}
\Robs(t;\nu)\ \equiv\ \frac{L_\nu^{\rm obs}(t)}{L_\nu^{\rm opt}(t)}
= \underbrace{\Rcore(t;\nu)}_{\rm compact}
+ \underbrace{\Rlobe(t;\nu)}_{\rm extended}
+ \underbrace{\Rsf(t;\nu)}_{\rm host}
+ \cdots\,.
\label{eq:Rsum}
\end{equation}
In practice $\Rsf$ can be estimated and subtracted using SED-based host SFRs and frequency-matched radio--SFR calibrations \citep[e.g.,][]{2018MNRAS.475.3010G}, with single-tracer estimates used cautiously \citep{1992ARA&A..30..575C,2001ApJ...554..803Y,2003ApJ...586..794B}. The jet-related loudness $\Rjet\equiv\Rcore+\Rlobe$ carries the physically relevant information, but $\Rsf$ is retained explicitly because it establishes a low-$R$ floor and can partially fill any intrinsic valley if left uncontrolled.

The compact and extended radio luminosities are parameterized as
\begin{align}
L_\nu^{\rm core}(t) &= A_\nu\,\epssyn^{\rm core}\,
                        \fbeam(\theta,\Gbulk)\,\Pjet(t)^{\beta_{\rm c}},
\label{eq:Lcore}\\[4pt]
L_\nu^{\rm lobe}(t) &= B_\nu\,\epssyn^{\rm lobe}\,\fesc(\Xienv)\,
                        \int_0^\infty \Knu(\tau)\,\Pjet(t-\tau)^{\beta_{\rm l}}\,d\tau\,.
\label{eq:Llobe}
\end{align}

The coefficients $A_\nu$ and $B_\nu$ absorb the mapping from kinetic power to monochromatic luminosity and may be calibrated against X-ray cavity measurements, inverse-Compton emission, or dynamical models \citep{1999MNRAS.309.1017W,2010ApJ...720.1066C,1997MNRAS.286..215K}. The factors $\epssyn^{\rm core,lobe}$ encode radiative efficiencies; $\beta_{\rm c}$ and $\beta_{\rm l}$ allow for non-linear power scalings. The beaming factor $\fbeam$ accounts for Doppler boosting \citep{1995PASP..107..803U}, and $\fesc$ captures whether launched jets break out to form large-scale lobes or dissipate within the nuclear region, as in GPS and CSS sources \citep{1998PASP..110..493O,2021A&ARv..29....3O,2012ApJ...760...77A,2025A&A...704A..93A}.

The function $\Knu(\tau)$ encodes synchrotron ageing, inverse-Compton losses, adiabatic expansion, and transport. For any individual source the detailed response need not be exponential. The minimal analytically tractable choice,
\begin{equation}
\Knu(\tau) = \frac{1}{\taunu}\exp\!\left(-\frac{\tau}{\taunu}\right),
\label{eq:exp_function}
\end{equation}
should be understood as an effective one-timescale approximation that retains a single fading time $\taunu$. It leads to the first-order evolution equation
\begin{equation}
\taunu\,\frac{d}{dt}L_\nu^{\rm lobe} + L_\nu^{\rm lobe}
= B_\nu\,\epssyn^{\rm lobe}\,\fesc\,\Pjet(t)^{\beta_{\rm l}}\,.
\label{eq:lobe_ode}
\end{equation}

To obtain the exact stationary solution quoted in the main text, we specialise to a two-state jet,
\begin{equation}
\Pjet(t)=P_0\,J(t), \qquad J(t)\in\{0,1\},
\end{equation}
where $J=1$ denotes a strong-jet episode and the weak state is assumed to contribute negligibly
to the large-scale radio source.
Because $J$ is binary, $\Pjet^{\beta_{\rm l}}=P_0^{\beta_{\rm l}}J$.
The reference extended-radio normalization of a permanently active source is then
\begin{equation}
L_{\nu,{\rm max}}^{\rm lobe}
= B_\nu\,\epssyn^{\rm lobe}\,\fesc\,P_0^{\beta_{\rm l}}\,,
\label{eq:Lmax_lobe}
\end{equation}
and the normalized variable $Y(t)\equiv L_\nu^{\rm lobe}(t)/L_{\nu,{\rm max}}^{\rm lobe}$ obeys Eq.~(\ref{eq:Y_ode_main}). The derivation concerns the radio numerator alone; the connection to $\Rlobe$ follows after division by the optical luminosity when that denominator is either slowly varying or controlled within the sample. For real evolving lobes, $L_{\nu,{\rm max}}^{\rm lobe}$ should be viewed as a normalization of this reference response, not as a claim that a physical lobe approaches a time-independent luminosity under constant jet power. Low scalar $R$ is not a single physical condition. It can reflect low instantaneous jet power, a low duty cycle, inefficient escape, host-emission dominance, or some combination thereof. Disentangling these requires the decomposed observables; scalar $R$ alone can not.

\section{Breakout time and the escape parameter}\label{app:escape}

The escape parameter is $\Xienv=t_{\rm on}/t_{\rm bo}$.
Here we relate the breakout time $t_{\rm bo}$ to standard momentum-balance estimates.

Consider a confining medium with characteristic scale height $H_{\rm ISM}$ and density
$\rho_a$.
If the jet head advances at speed $v_h$, then
\begin{equation}
t_{\rm bo}\ \simeq\ \frac{H_{\rm ISM}}{v_h}\,.
\label{eq:tbo_def}
\end{equation}
Balancing jet momentum flux $F_{\rm jet}\sim L_{\rm jet}/c$ against the ambient ram pressure
$\rho_a v_h^2 A_{\rm jet}$ gives
\begin{equation}
v_h\ \sim\ \left(\frac{L_{\rm jet}}{\rho_a c\,A_{\rm jet}}\right)^{1/2},
\label{eq:vh_scaling}
\end{equation}
up to order-unity factors depending on collimation and entrainment.
Combining these expressions, the breakout time increases with ambient density and decreases
with jet power:
\begin{equation}
\Xienv\ \equiv\ \frac{t_{\rm on}}{t_{\rm bo}}
\ \sim\ \frac{t_{\rm on}\,c}{H_{\rm ISM}}
  \left(\frac{\Pjet}{\rho_{\rm a}c^3 A_{\rm j}}\right)^{1/2}.
\label{eq:Xi_env}
\end{equation}
The smooth escape fraction of Eq.~(\ref{eq:fesc_eq}) encompasses both jets genuinely
frustrated by high ambient density and jets launched during episodes too short to achieve
escape, regardless of the ambient conditions.

\section{Stationary distribution in the two-state limit}\label{app:beta_deriv}

This appendix derives the stationary distribution used in Section~\ref{subsec:analytic_solution}.
For the two-state jet and exponential response of Appendix~\ref{app:function}, the normalized
extended-radio variable $Y$ obeys
\begin{equation}
\taunu\,\frac{dY}{dt} = J(t) - Y(t),
\end{equation}
with $J(t)\in\{0,1\}$ a telegraph process with transition rates $\lambda_\uparrow$ (0$\to$1) and $\lambda_\downarrow$ (1$\to$0). 

Let $p_0(y)$ and $p_1(y)$ be the stationary joint densities of finding $Y=y$ while the system is in states $J=0$ and $J=1$, respectively. Because the dynamics in each state is deterministic, these functions satisfy the stationary transport equations
\begin{align}
0 &= -\frac{d}{dy}\!\left[-\frac{y}{\taunu}p_0(y)\right]
     - \lambda_\uparrow p_0(y) + \lambda_\downarrow p_1(y), \label{eq:fp0}\\[4pt]
0 &= -\frac{d}{dy}\!\left[\frac{1-y}{\taunu}p_1(y)\right]
     + \lambda_\uparrow p_0(y) - \lambda_\downarrow p_1(y). \label{eq:fp1}
\end{align}
Adding Eqs.~(\ref{eq:fp0}) and (\ref{eq:fp1}) shows that the net probability flux in $y$ is
constant.
Since the support is bounded to $0<y<1$, a stationary solution requires that this flux vanish.
One therefore obtains
\begin{equation}
-y\,p_0(y) + (1-y)\,p_1(y) = 0
\quad\Rightarrow\quad
p_1(y) = \frac{y}{1-y}\,p_0(y)\,.
\label{eq:p1p0}
\end{equation}

Substituting Eq.~(\ref{eq:p1p0}) into Eq.~(\ref{eq:fp0}) gives a first-order ordinary
differential equation whose solution is
\begin{equation}
p_0(y)= C\,y^{\alpha-1}(1-y)^{\beta},\qquad
p_1(y)= C\,y^{\alpha}(1-y)^{\beta-1},
\label{eq:p0p1_exact}
\end{equation}
where
\begin{equation}
\alpha=\lambda_\uparrow\taunu,\qquad
\beta =\lambda_\downarrow\taunu,
\end{equation}
and $C$ is a normalisation constant determined by requiring $\int_0^1[p_0(y)+p_1(y)]\,dy=1$.
The integrals of $p_0$ and $p_1$ give the stationary probabilities of the two states:
\begin{equation}
\int_0^1 p_0(y)\,dy = \frac{\beta}{\alpha+\beta}\,,\qquad
\int_0^1 p_1(y)\,dy = \frac{\alpha}{\alpha+\beta}\,,
\end{equation}
which can be verified using standard Beta-function identities. The duty cycle is then
\begin{equation}
\fduty = \frac{\int_0^1 p_1(y)\,dy}{\int_0^1 p_1(y)\,dy + \int_0^1 p_0(y)\,dy} = \frac{\alpha}{\alpha + \beta} = \frac{\lambda_\uparrow}{\lambda_\uparrow + \lambda_\downarrow}\,.
\end{equation}

The marginal stationary density is therefore
\begin{equation}
p(y)=p_0(y)+p_1(y)
=\frac{1}{B(\alpha,\beta)}\,y^{\alpha-1}(1-y)^{\beta-1},
\end{equation}
which is the Beta distribution quoted in Eq.~(\ref{eq:beta_dist}).

The conditional distributions follow immediately:
\begin{equation}
p(y|J=1)=\frac{p_1(y)}{\fduty}\sim{\rm Beta}(\alpha+1,\beta),\qquad
p(y|J=0)=\frac{p_0(y)}{1-\fduty}\sim{\rm Beta}(\alpha,\beta+1),
\end{equation}
recovering Eq.~(\ref{eq:condY}). These conditional forms yield closed expressions for the observable fractions of recently triggered and remnant systems. For a survey threshold $y_{\rm th}$, the triggering fraction is $f_{\rm trig}(y_{\rm th})=\fduty\,I_{y_{\rm th}}(\alpha+1,\beta)$ and the remnant fraction is $f_{\rm rem}(y_{\rm th})=(1-\fduty)[1-I_{y_{\rm th}}(\alpha,\beta+1)]$, where $I_x(a,b)\equiv B(a,b)^{-1}\int_0^x t^{a-1}(1-t)^{b-1}dt$ is the regularized incomplete beta function.

These expressions are exact within the two-state, exponential-response limit. 
Thus the two-state, exponential-response limit gives a closed mapping between observables and model parameters: the duty cycle sets the mean; the memory-to-switching ratio sets the spread.

\section{Population modelling: from duty cycles to observed loudness distributions}\label{app:forward}

This appendix outlines the minimal population model connecting the latent engine parameters to an observed radio-loudness distribution and identifies where survey selection enters. The population calculation uses three inputs: (i) switching rates $(\lambda_\uparrow,\lambda_\downarrow)$, or equivalently $\fduty$ and $t_{\rm switch}$; (ii) a memory timescale $\taunu(\nu)$ for the lobe component; (iii) an escape fraction $\fesc(\Xienv)$ controlling whether a given ON episode contributes to extended emission. In the exponential fading limit, the Beta distribution (Eq.~\ref{eq:beta_dist}) provides the corresponding distribution of $Y$ directly.

The observed radio loudness is obtained by applying the survey weights of Eq.~(\ref{eq:radio_decomp}) to $(\Rcore,\Rlobe)$. The coefficients $w_{\rm c},w_{\rm l},w_{\rm SF}\in[0,1]$ encode detection efficiency for each component: high-resolution GHz surveys may have $w_{\rm l}\ll1$ because diffuse lobes are resolved out, while low-frequency arrays may have $w_{\rm c}\ll1$ if compact cores are confusion-limited. Star-formation contamination enters through $f_\nu^{\rm SF}$ (equivalently $\Rsf$), which can be modelled as a host-associated floor with scatter and constrained using independent SFR tracers. Comparing samples at different redshifts requires consistent K-corrections to convert between $\Robs$ and $\Rrest$ via Eq.~(\ref{eq:Kcorr}). Non-detections are not missing data: upper limits on $\Rlobe$ constrain the jet-off state and inform $\lambda_\uparrow$ and $\lambda_\downarrow$. Any inference of duty cycles from survey catalogues must incorporate censoring and the selection weights $w_i$ explicitly; neglecting them biases the recovered duty-cycle distribution toward detectable sources and systematically overestimates $\fduty$.

\clearpage

\clearpage

\bibliography{sample701}{}
\bibliographystyle{aasjournalv7}

\end{document}